# Competing soft phonon modes at the charge-density-wave transitions in DyTe3

*M. Maschek[1], D. A. Zocco[1,†], S. Rosenkranz[2], R. Heid[1], A. H. Said[3], A. Alatas[3], P. Walmsley[4], I. R. Fisher[4,5], F. Weber[1]*

[1] *Institute for Solid State Physics, Karlsruhe Institute of Technology, D-76021 Karlsruhe, Germany*
[2] *Materials Science Division, Argonne National Laboratory, Argonne, Illinois, 60439, USA*
[3] *Advanced Photon Source, Argonne National Laboratory, Argonne, Illinois, 60439, USA*
[4] *Geballe Laboratory for Advanced Physics and Department of Applied Physics, Stanford University, CA 94305*
[5] *The Stanford Institute for Materials and Energy Sciences, SLAC National Accelerator Laboratory, Menlo Park, CA 94025, USA*

[†] Present address: *Institute of Solid State Physics, Vienna University of Technology, 1040 Vienna, Austria*

**The family of rare-earth tritellurides RTe3 features charge-density-wave (CDW) order related to strongly momentum-dependent electron-phonon coupling. Similar to other CDW compounds, superconductivity is observed when the CDW order is suppressed via hydrostatic pressure [1]. What sets the heavier members of the RTe3 series apart is the observation of a second CDW transition at lower temperatures having an in-plane ordering wavevector $q_{CDW,2} \parallel [100]$ of almost the same magnitude but orthogonal to the ordering wavevector $q_{CDW,1} \parallel [001]$ observed at higher temperatures [2]. Here, we report an inelastic x-ray scattering investigation of the lattice dynamics of DyTe3. In particular, we show that there are several phonon modes along both in-plane directions, which respond to the onset of the CDW transition at $T_{CDW,1} = 308$ K. Surprisingly, these soft modes close to $q_{CDW,2} = (0.68, 0, 0)$ show strong softening near $T_{CDW,1}$ but do not exhibit any response to the lower-temperature transition at $T_{CDW,2} = 68$ K. Our results indicate that the low-temperature CDW order is not just the 90° rotated analogue of the one appearing at high temperatures.**

## I. Introduction

The research on charge-density waves (CDW), a modulation of the electronic density of states, intensified since the discovery of the CDW phase competing with superconductivity in copper-oxide superconductors [3-6]. Furthermore, it is now evident that many well-known CDW materials feature superconducting phases when the CDW phase is suppressed by extrinsic parameters such as pressure or intercalation in layered materials [1,7-11]. Investigating the mechanism of CDW formation itself, recent work [12-14] showed that CDW order is not generally be related to Fermi surface nesting as proposed in the seminal work of Peierls [15]. Instead, the full momentum dependence of both the electronic band structure and the electron-phonon coupling (EPC) matrix element have to be taken into account to explain the formation of CDW order and, hence, also that of superconductivity when CDW order is suppressed.

The quasi two-dimensional (2D) rare-earth tritellurides RTe3 (R = La-Nd, Sm, Gd-Tm) are prime examples of CDW ordered ground states. Angle-resolved photoemission spectroscopy (ARPES) measurements [16] [17] of several rare-earth tritelluride compounds lead to the assumption of a nesting driven CDW order and, hence, RTe3 are often regarded as canonical Peierls-like CDW systems. On the other hand, theoretical investigations of RTe3 [18] revealed a nesting feature, i.e. a peak in the imaginary part of the electronic susceptibility $\chi''(\mathbf{q})$, at a wave vector $\mathbf{q}_{FS}$ different from the observed CDW ordering wave vector $\mathbf{q}_{CDW}$. Nonetheless, the real and imaginary parts of $\chi(q)$ in those calculations show a broad enhancement in the correct direction but the CDW ordering wave vector was not correctly predicted [$\mathbf{q}_{FS} = (0,0,0.25)$ compared to the observed ordering wave vector $\mathbf{q}_{CDW} = (0,0,0.295)$][19]. More importantly, as the authors pointed out, the only modest enhancement of the susceptibility, far from a divergent-like behavior expected in a nesting scenario, is insufficient to drive the CDW instability. Theoretical analysis of Raman spectroscopy measurements [20] in ErTe3 further corroborated that the weak nesting geometry is not identical to the ordering wave vector and enhanced EPC matrix elements must be important.

In our previous investigation of the lattice dynamical properties of TbTe3 [14], we identified the soft phonon mode of the CDW transition at high temperatures [$T_{CDW,1}$ (TbTe3) = 330 K] and demonstrated the decisive impact of the momentum-dependent EPC matrix element for the CDW ordering wavevector $q_{CDW}$ by a combined inelastic x-ray and density-functional theory investigation. Here, we extend this work to DyTe3, which is well-known to display a first CDW transition at $T_{CDW,1} = 308$ K and a second one at $T_{CDW,2} \approx 50$ K [2]. Hence, the ground state of DyTe3 is characterized by the coexistence of two distortions with orthogonal ordering wavevectors, i.e. $q_{CDW,1} \approx (0,0,0.298)$ and $q_{CDW,2} \approx (0.656,0,0)$ where temperature dependent values were observed for $q_{CDW,1}$ as well as $q_{CDW,2}$ [21] similar to observations in, e.g., TbTe3 [2]. Full refinements of the atomic structure in the structurally distorted phases are lacking. Nonetheless, it is expected that both distortions are closely related as they appear both in the nearly tetragonal *ac*-plane in this layered material ($b \approx 6 \times a$). It is noteworthy that chemical pressure, i.e., by exchanging the R ion [2], and hydrostatic pressure [1,9] affect the CDW phase transitions in a similar manner: $T_{CDW,1}$ decreases and $T_{CDW,2}$ increases for increasing chemical/hydrostatic pressure.

In our measurements, we find indeed a strong softening of phonon modes along both in-plane directions with an energy difference of only 2 meV at $T_{CDW,1}$. However, the phonon modes at $q_{CDW,2}$ that exhibit a strong softening at $T_{CDW,1}$ show no measurable response to the phase transition at $T_{CDW,2}$. This indicates that the low-temperatures CDW transition is not the 90°-rotated analogue of the high-temperature transition and poses new questions about the observed phase competition of the two structural distortions, e.g., as function of pressure [9].

## II. Theory

We performed *ab-initio* calculations for the lattice dynamical properties based on *density-functional-perturbation-theory* (DFPT) using the high-temperature orthorhombic structure present at $T > T_{CDW,1}$. In RTe$_3$, the *f* states of the R ion are localized and shifted away from the Fermi energy. Therefore, they are expected to play no role in the physics of the CDW [17] [22]. To avoid complications with the *f* states in standard density-functional theory, we have performed our calculations for LaTe$_3$ in the framework of the mixed basis pseudopotential method [23]. The exchange-correlation functional was treated in the local-density approximation (LDA). Norm-conserving pseudopotentials were constructed following the scheme of Vanderbilt including *5s* and *5p* semicore states of La in the valence space but excluding explicitly *4f* states [24]. The basis set consisted of plane waves up to 20 Ry complemented with local functions of s, p, and d symmetry at the La sites. For the exchange-correlation functional the local-density approximation (LDA) was applied [25]. DFPT as implemented in the mixed basis pseudopotential method [26] was used to calculate phonon energies and electron-phonon coupling (EPC). An orthorhombic $24 \times 8 \times 24$ **k**-point mesh was employed in the phonon calculation, whereas an even denser $48 \times 12 \times 48$ mesh was used in the calculation of phonon line widths to ensure proper convergence. This was combined with a standard smearing technique using a Gaussian broadening of 0.2 eV.

## III. Experimental

The experiment was performed at the XOR 3-ID and 30-ID high energy-resolution inelastic x-ray scattering (HERIX) beamlines [27] [28] of the Advanced Photon Source, Argonne National Laboratory. The samples were high-quality single crystals of DyTe$_3$, grown by slow-cooling a self-flux with CDW transitions at $T_{CDW,1} = 308$ K and $T_{CDW,2} = 68$ K determined from elastic intensities in our investigation and in reasonable agreement with previous reports [2]. Samples had dimensions of $1 \times 1 \times 0.1 \text{mm}^3$ and were mounted in closed-cycle refrigerators allowing us to investigate a wide temperature range $10 \text{ K} \leq T \leq 400 \text{ K}$. In order to verify that our samples are not twinned, we measured elastic intensities at wavevectors $Q = (0,6,1)$ and $(1,6,0)$ of which only the former corresponds to an allowed Bragg reflection. The components ($Q_h$, $Q_k$, $Q_l$) of the scattering vector are expressed in reciprocal lattice units (r.l.u.) ($Q_h$, $Q_k$, $Q_l$) = (h*2π/a, k*2π/b, l*2π/c) with the lattice constants $a = 4.407$ Å, $b = 26.313$ Å and $c = 4.42$ Å of the orthorhombic unit cell. Measurements at small wavevector, i.e., $Q = (1,1,l)$ and $(1,0,l)$ were done on the HERIX spectrometer at 3-ID, which has a limited scattering angle of $2\Theta_{max} = 17°$. Energy scans at larger wavevectors were done on the HERIX spectrometer at 30-ID with $2\Theta_{max} = 31°$. The energy resolutions were 1.55 meV at 30-ID and 2.0 meV at 3-ID with incident energies of 23.78 keV (30-ID) and 21.66 keV (3-ID) respectively. Phonon excitations in constant *Q*-scans were approximated by damped harmonic oscillator (DHO) functions [29]. We extracted the intrinsic phonon line width Γ by fitting the resolution broadened DHO function to the experimental data. Thus, we obtain phonon energies $\omega_q = \sqrt{\widetilde{\omega}_q^2 - \Gamma^2}$, where $\widetilde{\omega}_q$ is the phonon energy renormalized only by the real part of the susceptibility $\chi$ and Γ is closely related to the imaginary part of $\chi$.

## IV. Results

### A. Theory

The unmodulated structure of quasi-2D RTe$_3$ is shown in Figure 1(a). The lattice is weakly orthorhombic and usually described based on an orthorhombic unit cell (space group $Cmcm$, #63). It consists of two structural motifs: nominally square planar Te sheets and corrugated RTe slabs. Note that in this space group setting, the *b* axis is oriented along the long crystal axis, perpendicular to the Te planes. It is instructive to add a note on the implications of the orthorhombic structure for momentum space *Q*. The orthorhombicity results in altered static structure factors for Bragg scattering compared to the tetragonal case with lattice constants $a_{tet} = c_{tet}$. While in the latter, the zone boundaries along both the [100] and [001] direction would be $q_{tet} = (0.5,0,0)$ and $(0,0,0.5)$, respectively, the zone boundaries in the orthorhombic $Cmcm$ structure are $q = (1,0,0)$ and $(0,0,0.5)$. Hence, the wavevectors $q = (0,0,0.3)$ and $(0,0,0.7)$ are equivalent but the wavevectors $q = (0.3,0,0)$ and $(0.7,0,0)$ are not. This is evident in the below presented calculations, e.g., in Figures 1(b)(c).

Before we discuss the results of our calculations, we note that we performed the phonon calculations on a denser momentum grid than those presented in our work on TbTe$_3$ [14]. The phonon energies calculated within DFPT at a particular wavevector can be different than the energy at that wave vector deduced from interpolating phonon energies calculated at neighboring wavevectors in a less dense momentum grid. Such differences can be expected to be particularly large in the vicinity of a structural instability because the interpolation does not account for anomalies in the electronic structure and EPC effects. Hence, the current calculations feature more pronounced imaginary energies than presented previously for TbTe$_3$ [shown as negative values in Figure 1(b)(c) for simplicity].

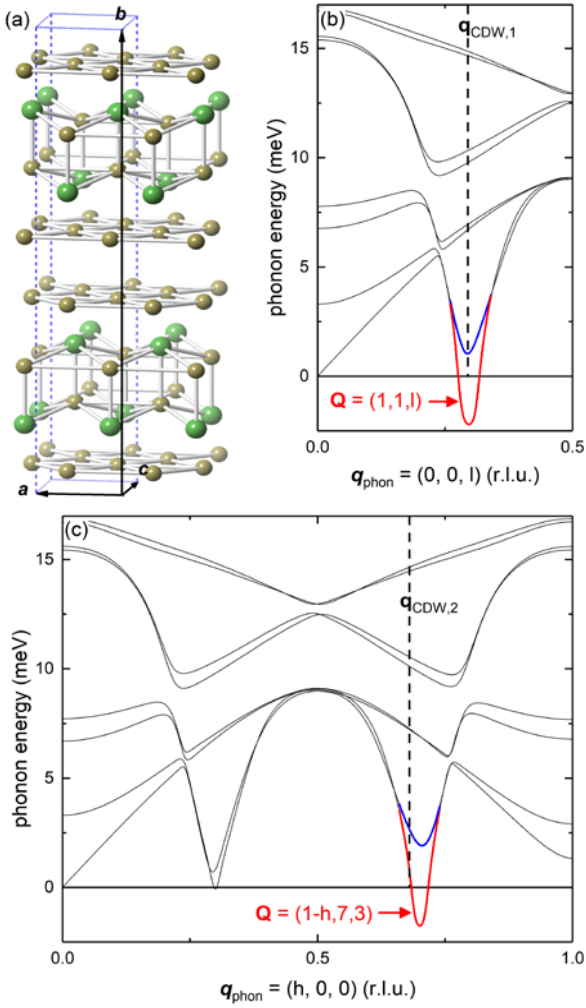

**FIG. 1.** (a) Orthorhombic unit cell of the crystalline structure of RTe$_3$ above $T_{CDW,1}$ [R: large (green) spheres; Te: small (yellow) spheres]. (b,c) Calculated dispersion for in-plane transversely polarized phonon modes along the (a) [001] and (b) [100] directions. The given $Q$ values denote the wave vectors at which the respective soft phonon modes (red line sections) have large structure factors, i.e., scattering intensities. Corresponding values for the second-lowest phonon modes at the respective wave vectors (blue line sections) are (b) $Q = (1,0,l)$ and (c) $Q = (h,0,4)$. The dashed vertical lines indicate the experimentally observed ordering wavevectors $\mathbf{q}_{CDW,1} = (0,0,0.295)$ and $\mathbf{q}_{CDW,2} = (0.32,0,0)$.

Figures 1(b)(c) show calculated phonon dispersions for LaTe$_3$ where the atomic mass of La was substituted by that of Dy in order to correct the phonon energies involving Dy displacements. Otherwise, we argue that $f$ electrons are not important and, hence, the simplified calculation for LaTe$_3$ can be compared with experimental results for members of the RTe$_3$ family such as TbTe$_3$ [14] and DyTe$_3$. We show the dispersions along $q = (0,0,l)$ [Fig. 1(b)] and $(h,0,0)$ [Fig. 1(c)] for $0 \leq l \leq 0.5$ and $0 \leq h \leq 1$, respectively, where the end points, i.e. $l = 0.5$ and $h = 1.0$, correspond to the Brillouin zone boundaries along the respective directions in momentum space. As we have shown in previous work, DFPT correctly predicts the softening of a transverse optic (TO) phonon dispersing along the [001] direction [thick red line section in Fig. 1(b)]. The detailed analysis of this CDW soft phonon mode was already the subject of a previous publication in which we presented IXS measurements for TbTe$_3$ [14]. There, we showed that observed intensities agree very well with the calculated structure factors and, hence, can confirm the TO character predicted by DFPT. We will see below that the behavior in DyTe$_3$ at $q_{CDW,1}$ is very similar to that observed in TbTe$_3$. However, there is a second phonon which exhibits an only slightly less strong anomaly in our calculation [thick blue line section in Fig. 1(b)]. In the following we will refer to the two phonon modes at $q_{CDW,1}$ showing pronounced minima in their respective dispersions as the leading soft phonon mode [red line in Fig. 1(b)] and second soft phonon mode [blue line in Fig. 1(b)] for simplicity although the latter does not soften completely at any temperature in DyTe$_3$. We use analogue expressions for the phonon modes at $q_{CDW,2}$ [see below and red/blue lines in Fig. 1(c)]. Structure factor calculations show that intensities of the leading soft mode propagating along the [001] direction are strongest at $Q = (3,7,0.3)$ and $(3,1,0.3)$ whereas the second soft mode is best measured close to $Q = (1,0,0.3)$. This might indicate that correlations in the lattice distortion along the out-of-plane axis are important to correctly predict the modulated CDW phase in RTe$_3$.

Along the orthogonal [100] direction we find two wave vectors, i.e., $q \approx (0.3,0,0)$ and $(0.7,0,0)$, each featuring two modes with pronounced dispersion minima. The calculated structure factors show the equivalent behavior as along the [001] direction, i.e., the structure factors of the leading soft modes at $(0.3,0,0)$ and $(0.7,0,0)$ are largest at $Q = (h,7,3)$ and $(h,1,3)$ and those of the second modes for wavevectors with $K = 0$, e.g., at $Q = (h,0,4)$. The anomaly in the lower-energy mode is more pronounced at $h = 0.7$ than at 0.3. In our experiment we focused on the dispersion at $0.5 \leq h \leq 0.85$ and indeed found strong CDW superlattice peaks at $h = 0.68$ for $T \leq T_{CDW,2}$ (see below and Fig. 3). Comparing the energies of all soft modes, at $q_{CDW,1} \approx (0,0,0.3)$ and $q_{CDW,2} \approx (0.7,0,0)$ [red line sections in Figs. 1(b) and (c)], we find that the structural instability is most pronounced at $q_{CDW,1}$. Hence, DFPT predicts correctly that the orthorhombic structure of RTe$_3$ should be modulated with a periodicity corresponding to $\mathbf{q}_{CDW,1}$. The corresponding atomic displacement patterns exhibit large Te movements along the respective transverse directions but only for Te atoms in the planar Te sheets. Movements for atoms in the corrugated RTe slabs show only small (Dy) or very small values (Te).

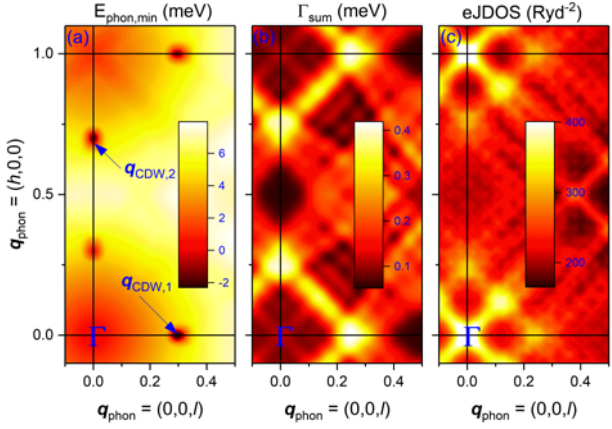

**FIG. 2.** Color-coded contour plots of the calculated (a) minimum phonon energy, (b) energy-integrated phonon linewidth and (c) electronic joint density of states (eJDOS) for wave vectos in the $(h, 0, l)$ plane. The CDW ordering wavevectors are indicated in (a).

Figure 2 provides an overview of the calculated lattice dynamical properties in the $(h, 0, l)$ plane: We show the energy of the lowest phonon mode (color code), the sum of the linewidths of all individual phonon modes $\Gamma_{sum}$ and the electronic joint density-of-states (eJDOS) for each phonon wavevector in the $\mathbf{q} = (h,0,l)$ plane [Figs. 2(a)-(c)]. Typically, the lowest energy for any given wavevector corresponds to an acoustic phonon mode. However, we can clearly observe four well-defined regions with anomalously low phonon energies where the characters of the above discussed TO soft modes dive below the normal acoustic modes. The calculated anomalies occur at wavevectors corresponding to the observed CDW ordering wavevectors $q_{CDW,1}$ and $q_{CDW,2}$. Slightly less pronounced features are visible at wavevectors which would be equivalent to $q_{CDW,1}$ and $q_{CDW,2}$ in the undistorted tetragonal structure, i.e., at $q = (1,0,0.3)$ and $q = (0.3,0,0)$ (see discussion above). The plot demonstrates that each anomaly is sharply localized in momentum space. In particular, our results show that there are no "valleys" in the phonon energy landscape connecting closeby anomalies, e.g., at $q = (0.3,0,0)$ and $(0,0,0.3)$.

There has been an intense discussion on the origin of phonon anomalies in CDW materials in recent years [12-14,18,20,30,31]. One of the driving questions was whether the momentum dependence of the electronic susceptibility $\chi_q$ or that of the EPC matrix elements $\eta_q$ determines $q_{CDW}$. This can be best illustrated via the criterion for a stable CDW phase with a modulation wavevector $\mathbf{q}$ derived by Chan and Heine as [32]

$$\frac{4\eta_q^2}{\hbar\omega_q} \geq \frac{1}{\chi_q} + (2U_q - V_q)$$

where $\eta_q$ is the electron-phonon coupling matrix element associated with a mode at an energy of $\omega_q$, $\chi_q$ is the dielectric response of the conduction electrons, and $U_q$ and $V_q$ are their Coulomb and exchange interactions. Obviously, the imbalance can be driven either by a maximum in $\chi_q$ and/or $\eta_q$ at the CDW ordering wavevector.

With regards to measurements of the CDW soft phonon mode, it is instructive to compare the wavevector dependences of the phonon energy $\omega_{q\lambda}$ and the electronic contribution to the phonon linewidth $\gamma_{q\lambda}$ given in our calculations by [33]

$$\gamma_{q\lambda} \propto \omega_{q\lambda} \sum_{kvv'} \left|\eta_{k+qv',kv}^{q\lambda}\right|^2 \delta(E_{kv} - E_F)\delta(E_{k+qv'} - E_F)$$

for the phonon mode $\lambda$ at the wavevector $\mathbf{q}$ with that of the eJDOS at wavevector $\mathbf{q}$. While we present corresponding results in comparison to our IXS measurements in the experimental section (Fig. 4), we provide the momentum dependence of $\Gamma_{sum}$ and the eJDOS in Figs. 2(b) and (c). It is evident that the soft phonon anomalies visible in Figure 2(a) occur at wavevectors with maxima in $\Gamma_{sum}$ [Fig. 2(b)] whereas the maxima in the eJDOS are located along the [101] directions, e.g., at $q \approx (0.125,0,0.125)$ [Fig. 2(c)]. It is interesting to note that there are streaks of large values of $\Gamma_{sum}$ along the line connecting the soft phonon anomalies, e.g., at $q \approx (0.3,0,0)$ and $(0,0,0.3)$ including $q \approx (0.125,0,0.125)$ where eJDOS is maximum. However, the calculations predict no corresponding soft phonon mode indicating that the EPC at these wavevectors is connected to higher-energy modes not involved in CDW formation.

Focusing on the behaviour at $q_{CDW,1} \approx (0,0,0.3)$, we find that the maximum in $\Gamma_{sum}$ corresponds to a local maximum in the eJDOS at $q = (0,0,0.25)$. However, the soft mode anomaly is clearly offset along the [001] direction by about 0.05 reciprocal lattice units. Therefore, our calculations demonstrate that RTe$_3$ exhibit an intricate interplay of momentum dependences originating from the electronic structure and from the momentum dependence of EPC matrix elements.

We already know that the leading soft phonon mode along the [001] direction [Fig. 1(b)] is indeed the soft phonon mode of the high-temperature CDW transition from our work in TbTe$_3$ [14]. Hence, the predicted simultaneous softening of analogous phonon modes along the orthogonal in-plane direction [Fig. 1(c)] naturally leads to the question whether these modes are related to the low-temperature CDW transition. Ideally, one would like to perform phonon calculations in the distorted structure corresponding to the experimental situation at $T_{CDW,2} < T < T_{CDW,1}$. However, while we were able to perform such calculations for the superstructure of the commensurate CDW compound $1T$-TiSe$_2$ [34], the required supercells for any approximation of the incommensurate CDW order in DyTe$_3$ are too large for comparable lattice dynamical calculations.

### B. Experiment

The prediction of competing soft phonon modes from DFPT combined with the question about the relation between high- and low-temperature CDW transitions motivated us to experimentally investigate the soft phonon modes in DyTe$_3$. Although the existence of the second CDW transition is a well-known fact, there is little known about the structural distortion below $T_{CDW,2}$. In total, we have performed three experiments at the Advanced Photon Source at Argonne National Laboratory: At sector 3-ID we focused on the temperature dependence of phonons at $q = q_{CDW,1}$ on cooling across the high-temperature CDW transition. At sector 30-ID we performed a first experiment on the momentum dependence of the dispersion along the [001] direction and a second one focusing on the [100] direction. In the following we will first discuss results from elastic scattering, i.e., measurements with the HERIX spectrometer set to zero energy transfer. Then we will discuss results connecting to our previous work on TbTe$_3$ regarding the momentum dependence of the dispersion and line width of the soft mode along the [001] direction for $T \approx T_{CDW,1}$. At the end of this section we will present the temperature dependences of the predicted soft phonon modes at $q_{CDW,1}$ and $q_{CDW,2}$.

#### 1. Elastic scattering

Figure 3(a) shows scattering intensities at the two orthogonal CDW ordering wavevectors indicating transition temperatures of $T_{CDW,1} = 308$ K and $T_{CDW,2} = 68$ K for the high- and low-temperature transitions, respectively. These measurements were done with low momentum resolution ( $\Delta Q = 0.066 \text{ Å}^{-1}$ ) typically employed in inelastic experiments. This setup allowed us to measure the temperature dependent intensities of the superlattice reflections essentially neglecting the small shift of the ordering wavevectors. The difference between our estimated $T_{CDW,2} = 68$ K and the value reported by resistivity measurements of about 50 K are not inconsistent in that the signature of the lower-temperature CDW transition is rather weak and broad in resistivity [2]. On the other hand, we identify $T_{CDW,2}$ by the increase of elastic scattering at $q_{CDW,2}$. The best way to determine $T_{CDW,2}$ would be to determine the temperature at which long-range order sets in. Because of our coarse momentum resolution we were not able observe changes in the linewidth of momentum scans at $E = 0$ [see Fig. 3(b)].

Figure 3(b) shows intensities at $E = 0$ along the [100] direction in the Brillouin zone adjacent to $\tau = (1,7,3)$ where we observe a strong superlattice peak below 68 K. Though barely, our data reveal a shift of the peak position on cooling from $T = 50$ K to 10 K in agreement with diffraction results [21].

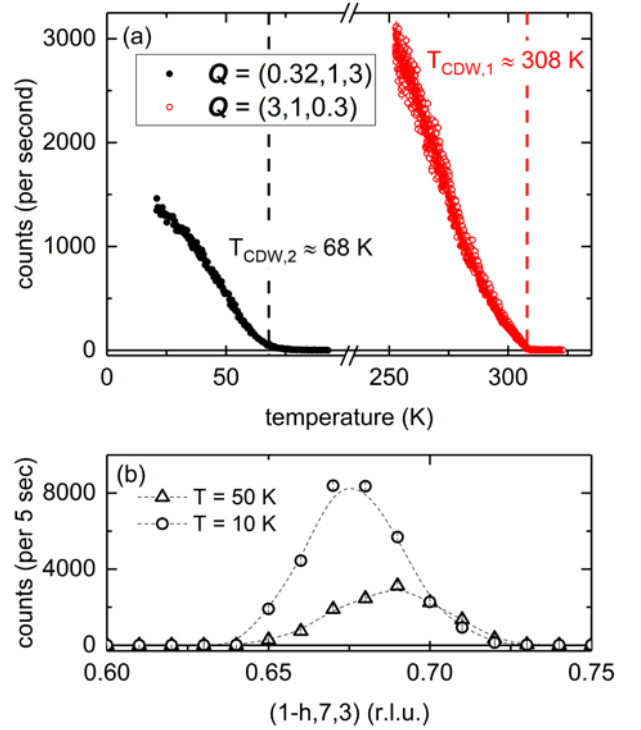

**FIG. 3.** (a) Observed elastic intensities at the in-plane ordering wavevectors $q_{CDW,1} \approx (0,0,0.3)$ and $q_{CDW,2} = (0.68,0,0)$ of the two different CDW orders in DyTe$_3$. Dashed vertical lines indicate the deduced transition temperatures $T_{CDW,1} \approx 308$ K and $T_{CDW,2} \approx 68$ K. (b) Scans for $E = 0$ along $q = (1-h,7,3)$ revealing a clear shift of $q_{CDW,2}$ on cooling from $T = 50$ K to 10 K.

#### 2. Phonon dispersion at $T \approx T_{CDW,1} (= 308$ K$)$

Figure 4 summarizes observed phonon dispersions (filled symbols) along the [001] [Fig. 4(a)] and [100] directions [Fig. 4(c)] together with the respective phonon linewidths [Figs. 4(b)(d)]. The data were taken in Brillouin zones adjacent to reciprocal wavevectors $\tau = (3,1,0)$ and $(1,7,3)$ where DFPT predicted large structure factors for the leading soft modes indicated by the thick red lines in Figures 1(b) and (c). Hence, we compare the experimental results with the corresponding dispersion lines featuring finite structure factors in the above given Brillouin zones and find overall a good agreement. As expected, we find also a close similarity with our previous results for TbTe$_3$ [14] with regard to both phonon dispersions and linewidths along the [001] direction [open green circles in Figs. 4(a) and (b)]. The calculation puts the minimum of the dispersion slightly above $l = 0.3$ r.l.u. whereas the experimental superlattice peak position is typically found slightly below $l = 0.3$ r.l.u. [2,21,35]. Regarding the minimum of the dispersion the agreement between experiment and calculation is even better along the [100] direction [Fig. 4(c)]. In particular, the wavevector position of the minimum of the calculated dispersion is well reproduced by experiment. However, the upward dispersion at $h < 0.7$ is steeper than predicted. As discussed previously, the soft mode has a TO character which starts dispersing from around 17 meV at the zone center and then bends down along both in-plane directions. The corresponding data points are marked in red in Figure 4. A comparison of the corresponding phonon linewidth with that observed for the high-energy optic mode dispersing always above 13 meV (blue symbols) reveals a strong momentum dependence, which is defined by an intricate interplay of the electronic band structure and the momentum dependent EPC matrix elements as shown in previous work [14]. Therefore, our experimental results at $T \approx T_{CDW,1}$ corroborate the prediction by DFPT that there is a strong competition of phonon modes in orthogonal directions of DyTe$_3$.

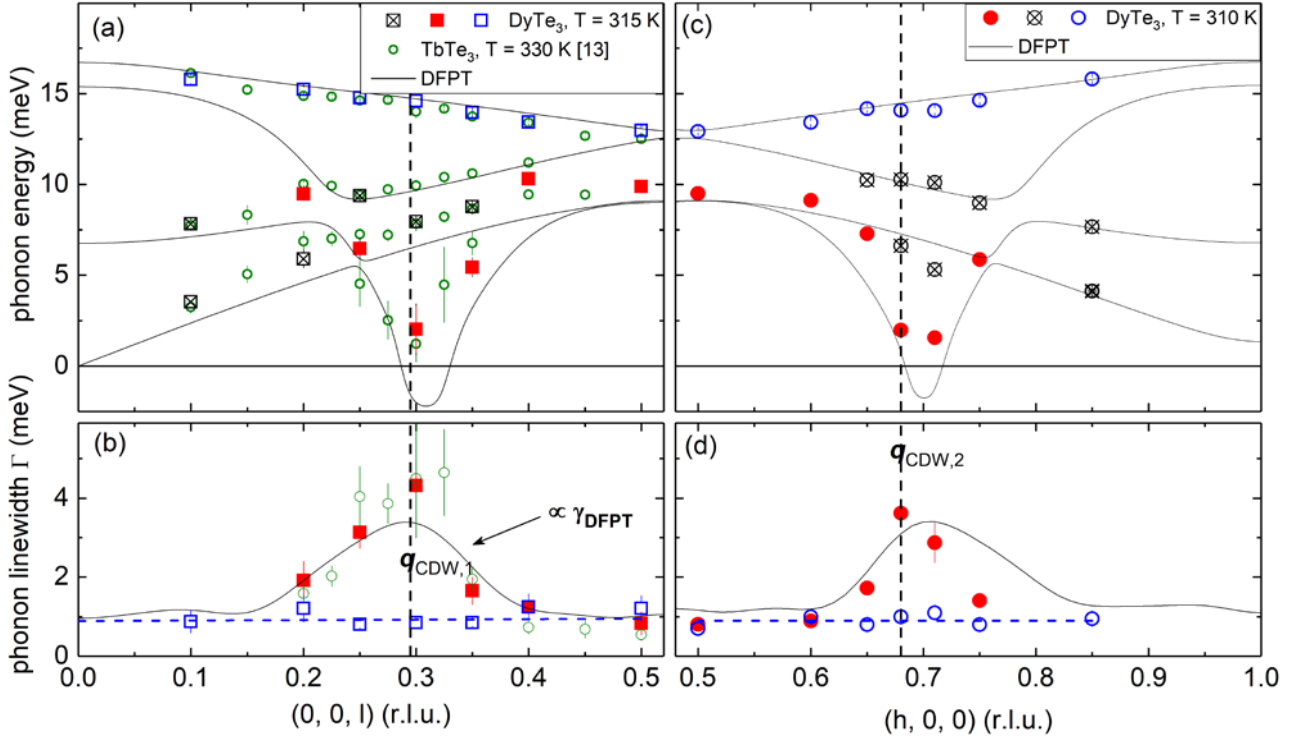

FIG. 4. (a) Comparison between calculated (solid lines) and observed (squares) energies of transversely polarized phonon modes propagating along the [001] direction in DyTe$_3$. Measurements were performed at $Q = (3,1,l)$, $l = 0.1 − 0.5$ and $T = 315$ K. Circles represent results from previous IXS measurements in TbTe$_3$ at $T = 330$ K taken from Ref. [14]. Red and blue squares mark the energies of the phonon modes for which the corresponding linewidths Γ are displayed in panel (b). (b) Wavevector dependences of linewidths Γ of the soft phonon mode (red squares) and optical mode (blue squares) at $T = 315$ K. The solid line represents a DFPT calculation of the linewidth of the soft mode. For comparison it has an offset (dashed blue line) and is scaled (see text). Green circles again represent published results for TbTe$_3$ [14]. Dashed vertical lines denote the ordering wave vector $q_{CDW,1}$ of the high-temperature transition. (c)(d) Analogue plots for the transversely polarized phonon mode propagating along the orthogonal [100] direction. Measurements were performed at $Q = (1 − h, 7, 3)$, $h = 0.5 − 0.85$ and $T = 310$ K. There are no corresponding results for other RTe$_3$ in the literature for this direction.

### 3. Temperature dependence of soft modes

Our calculations predict that there are several phonon modes along each in-plane direction displaying anomalies near the two orthogonal CDW ordering wavevectors [Fig. 1(b)(c)]. In an experiment at sector 3, we focused on the two modes dispersing along the [001] direction [Fig. 1(b)]. According to DFPT the leading (second) soft mode should acquire a large (negligible) structure factor at $Q = (1,1,0.3)$ and a negligible (large) structure factor at $Q = (1,0,0.3)$. Corresponding raw data are shown in Figures 5(a) and (c) for $T = 400$ K. Indeed, we observe a strong phonon peak at low energies of 4-5 meV in each scan. However, the fact that we see different modes only becomes evident when we cool down to $T = 310$ K $\approx T_{CDW,1}$ [Figs. 5(b),(d)]: The soft phonon at $Q = (1,1,0.3)$ [Fig. 5(b)] softens to about 1 meV and is almost critically damped. Furthermore, a resolution-limited elastic line develops [dash-dotted line in Fig. 5(b)] indicating the onset of the static CDW superstructure. On the other hand, the second soft phonon mode with a large intensity at $Q = (1,0,0.3)$ [Fig. 5(d)] softens as well but clearly remains at a finite energy of about 3 meV and no superlattice peak develops. This result also agrees with the typical expectation that strong superlattice peaks develop in Brillouin zones in which the soft phonon mode of the structural phase transition has a large dynamic structure factor above the transition temperature, which then freezes into the static structure factor of the elastic scattering on cooling through the transition temperature. Reversely, we do not expect to see any CDW peak at $Q = (1,0,0.3)$ since the leading soft mode has practically zero intensity – in agreement with experiment [Fig. 5(d)]. Finally, we note that measurements at $Q = (1,1,0.3)$ are not possible at $T < T_{CDW,1}$ due to the presence of the CDW superlattice peak. The intrinsic resolution function of IXS features a strong Lorentzian tail and, hence, a strong elastic signal makes phonon measurements at zone centers - $q_{CDW,1}$ becomes a zone center at $T < T_{CDW,1}$ - impossible.

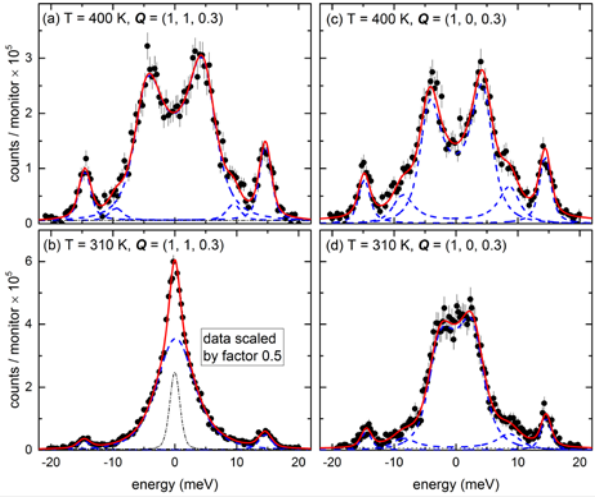
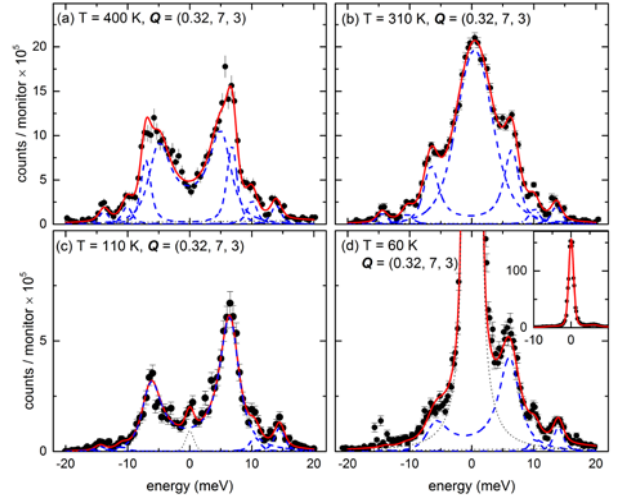

**FIG. 5.** Energy scans taken at (a)(b) $Q = (1, 1, 0.3)$ and (c)(d) $(1,0,0.3)$ for (top) $T = 400$ K and (bottom) 310 K, i.e., well-above and very close to $T_{CDW,1} = 308$ K. Solid (red) lines are fits consisting of damped harmonic oscillators (inelastic, blue dashed lines), estimated background (straight dotted line) and a pseudo-Voigt function for the elastic peak in (c) (black dash-dotted line). The observed intensities in (b) are scaled by a factor of 0.5 in order to plot the data at the same temperature on the same scale.

**FIG. 6.** Energy scans taken at $Q = (0.32,7,3)$ and different temperatures (a) $T = 400$ K, (b) 310 K, (c) 110 K and (d) 60 K. Solid (red) lines are fits consisting of damped harmonic oscillators (inelastic, blue dashed lines), estimated background (straight dotted line) and a pseudo-Voigt function for the elastic peak in (c) and (d) (black dotted line). The inset in (d) shows the data taken at $T = 60$ K but on a larger vertical scale in order to include the maximum intensities at zero energy transfer.

We next focus on the temperature evolution of two phonon modes for which theory predicts dispersion anomalies at $q_{CDW,2}$. We already discussed the observed dispersion at $Q = (1 - h, 7, 3)$ at $T \approx T_{CDW,1}$ [Fig. 4(c)] and show raw data taken at $Q = (0.32,7,3)$ at selected temperatures in Figure 6. The data at $T = 310$ K $\approx T_{CDW,1}$ reveal four different phonon modes in agreement with our structure factor calculations for this wavevector where the second soft mode should have essentially zero intensity [Fig. 6(b)]. Here, the finite energy of the lowest-energy peak cannot be resolved graphically anymore. We observe strong hardening of this mode both on cooling below and heating above $T_{CDW,1}$. While we can still distinguish four modes at 400 K [Fig. 6(a)], the overlap with the second lowest mode becomes very large below $T = 120$ K and we observe only three peaks at $T = 110$ K [Fig. 6(c)]. Finally, we present data at $T = 60$ K, i.e., just below the second CDW transition temperature $T_{CDW,2} = 68$ K [Fig. 6(d)]: While we clearly observe a strong superlattice peak at $Q = (0.32,7,3)$, we do not observe any soft mode behavior similar to what we observe at the high-temperature transition.

Figure 7 displays IXS data taken at a wavevector for which DFPT predicts large structure factors for the second soft mode at $q_{CDW,2}$, i.e. $Q = (0.68,0,4)$, and negligible intensities for the leading soft mode. While the predicted large intensity for the second soft mode is verified by experiment, the actual phonon energy is always much larger than the predicted 1.9 meV. Further, any softening and broadening are comparatively small and become only visible as function of temperature after detailed analysis and will be discussed below.

Figure 8 shows the temperature dependences of the energies and linewidths of the leading and second soft phonon modes predicted to show pronounced dispersion anomalies at $q_{CDW,1}$ and $q_{CDW,2}$ [see Fig. 1(b)(c)]. These are the modes discussed above based on the raw data shown in Figures 5, 6 and 7. To summarize, we observe signatures of the CDW transition at $T_{CDW,1} = 308$ K in all four phonon modes predicted to show dispersion anomalies by DFPT. Phonon intensities are consistent with the structure factors based on the calculated atomic displacement patterns. The leading instability is correctly calculated at $q_{CDW,1}$. However, the softening and broadening effects in the leading soft mode at $q_{CDW,2}$ are only slightly smaller in agreement with theory. While we observe a response of the second soft modes at $q_{CDW,1}$ and $q_{CDW,2}$ to the onset of the CDW phase at $T_{CDW,1}$, the effects are much smaller than in the leading soft modes. In particular, the energy of the second soft mode at $q_{CDW,2}$ is more than three times larger than calculated [Fig. 8(b)].

Yet, the most surprising result is the absence of any response of the leading and second soft modes at $q_{CDW,2}$ to the onset of the low temperature CDW phase on crossing $T_{CDW,2} = 68$ K. This is even more puzzling regarding the observation of a strong elastic superlattice peak at $Q = (0.32,7,3)$, i.e., the wavevector, where the leading soft mode at $q_{CDW,2}$ is strong – in theory and experiment.

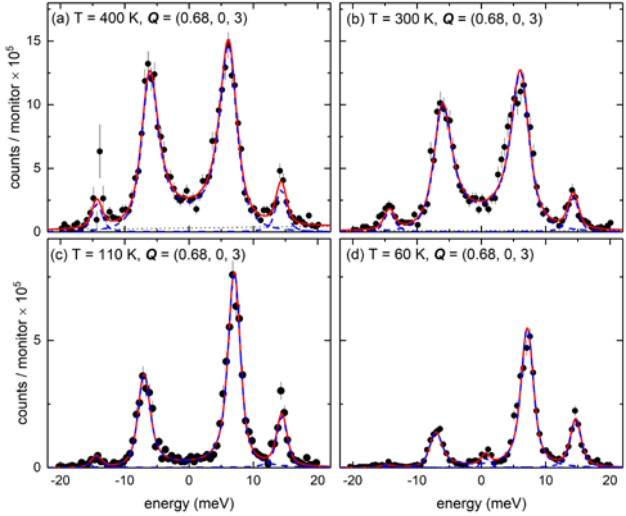

**FIG. 7.** Energy scans taken at $Q = (0.68,0,4)$ and different temperatures (a) $T = 400$ K, (b) 300 K, (c) 110 K and (d) 60 K. Solid (red) lines are fits consisting of damped harmonic oscillators (inelastic, blue dashed lines), estimated background (straight dotted line) and a pseudo-Voigt function for the elastic peak in (c) and (d) (black dotted line).

## V. Discussion

Our experimental and theoretical results on lattice dynamics in DyTe$_3$ reveal two contrasting pictures: (1) We find good agreement between *ab-initio* calculations and phonon spectroscopy for the CDW transition at $T_{CDW,1}$= 308 K. (2) We do not observe any phonon softening and/or broadening on cooling to and below $T_{CDW,2} = 68$ K – even though we observe a continuously rising strong superlattice peak.

Regarding the high-temperature transition we discuss our results with respect to time-resolved pump-probe studies using reflectivity [36], resonant soft x-ray diffraction [37] and angle-resolved photoemission spectroscopy [38,39]. From the time dependence of the different experimental observations, the authors could deduce characteristic phonon frequencies of about 1.7-1.75 THz and 2.2-2.4 THz for (Tb/Dy)Te$_3$ [36,38,39]. Results for CeTe$_3$ [38] reveal larger frequencies of 2.2 THz and 2.7 THz as can be expected because of the higher CDW transition temperature [40]. All of these measurements are sensitive to the amplitude mode in the ordered phase, i.e., below $T_{CDW,1}$. Hence, our data at $q_{CDW,1}$ and above $T_{CDW,1}$ are complementary as can be seen in a comparison with the detailed temperature dependence obtained via time-resolved reflectivity in DyTe$_3$ for T ≤ 260 K (Fig. 9) [36]. Following the reported intensity distribution, we can indicate the likely temperature dependence of the CDW amplitude mode which corresponds to the soft phonon mode at $T > T_{CDW,1}$ [(red) line in Fig. 9. Aside from the soft mode itself, our DFPT calculations predict phonons with the same symmetry at 6.4 meV and 9.6 meV [thick horizontal bars at $T \approx T_{CDW,1}$ in Fig. 9, see also Fig. 4(a) at $q \approx q_{CDW,1}$]. Considering the typical hardening of phonon energies on cooling, these two modes likely correspond to the low-temperature phonons observed by time-resolved reflectivity at 7 meV and 10.5 meV. It is well-known that phonon branches of the same symmetry are not allowed to cross but undergo an exchange of eigenvector, also called an anti-crossing. A similar thing can happen when a phonon mode displays a pronounced hardening (or softening) and thereby "crosses" energies of other phonon modes of the same symmetry. On cooling to low temperatures, the character of the amplitude/soft mode in DyTe$_3$ is probably being distributed among symmetry-related branches and that is the reason why time resolved measurements observe multiple amplitude modes.

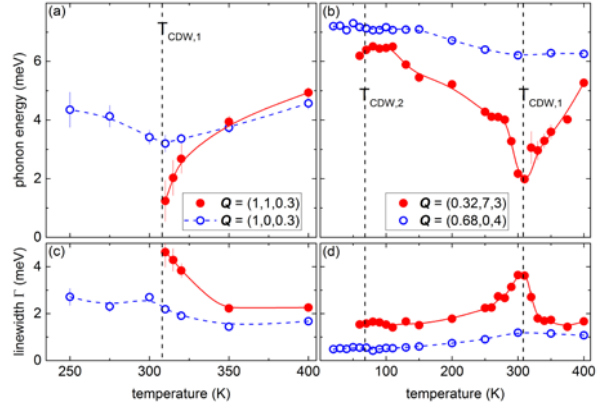

**FIG. 8.** Temperature dependent (a)(b) energies and (c)(d) linewidths of the predicted soft phonon modes at the CDW ordering wavevectors (a)(c) $q_{CDW,1} \approx (0,0,0.3)$ and (b)(d) $q_{CDW,2} = (0.68,0,0)$. Dashed vertical lines denote the respective CDW transition temperatures. Lines are guides to the eye. Absolute wavevectors for the measurements are given in the legends.

Now, we turn to the CDW transition at $T_{CDW,2} \approx 68$ K and the notable absence of any soft mode behaviour in our measurements. It could be argued that IXS is of course a highly momentum-selective method. Measurements might have been performed in a Brillouin zone not suited to observe the soft mode of the low-temperature CDW. However, the usual idea of a second-order structural phase transition, which takes place in DyTe$_3$ at $T_{CDW,2} = 68$ K [see Fig. 3(a)], is that a phonon mode softens to zero energy. At that point the dynamical structure factor of the phonon is transferred (at least partially) to the static structure factor of the superlattice in the ordered phase. Hence, one should observe the soft mode at wavevectors, displaying strong superlattice reflections in the ordered phase. We do find such a strong CDW peak at $Q = (0.32,7,3)$ [Fig. 6(d)], yet no soft mode behaviour is found [Fig. 8(b),(d)]. A second argument against our measurements missing the soft mode is that the absence of an amplitude mode below $T_{CDW,2}$ was also noted in the above discussed time-resolved reflectivity measurements [36].

However, CDW transitions without a soft phonon mode have been reported for a number of materials, e.g., NbSe$_3$

[41], (TaSe$_4$)$_2$I [42], $2H$-TaSe$_2$ [43], Nb$_3$Sn [44], ZrTe3 [45] and, most recently, YBa$_2$Cu$_3$O$_{6.6}$ [4]. Common to all these materials is that a resolution-limited "central" peak appears at the elastic position without a phonon softening to zero energy, which it is understood as defect-induced nucleation of finite-size domains of the low-temperature phase. While in general the nature of the lattice defects responsible for the central peak in these materials has remained undetermined, a model in the strong coupling regime for quasi 1D material, e.g., NbSe$_3$ and (TaSe$_4$)$_2$I, has been proposed by Aubry, Abramovici and Raimbault [46] and was used to explain experimental results from inelastic x-ray scattering in these materials [47,48].

A close inspection of the lattice-dynamical properties of the above quoted materials, however, reveals the following: while the softening of a particular phonon mode seems to be indeed incomplete when elastic scattering sets in, all of the above given compounds display signatures of the CDW phase transition in their lattice dynamical properties. The signatures range from partial phonon softening near $T_{CDW}$ [TaSe$_2$, (TaSe$_4$)$_2$I], strongly increased linewidth of a phonon mode at $T_{CDW}$ (NbSe$_3$) to complete softening but only at the transition temperature whereas the elastic scattering already sets in above (Nb$_3$Sn, ZrTe$_3$). YBa$_2$Cu$_3$O$_{6.6}$ features broad phonons as well. However, the situation is complicated by the presence of superconductivity having a strong impact on the anomalous phonon mode [4]. Hence, DyTe$_3$ is set apart from these examples in that we do not observe phonon softening or broadening at $\boldsymbol{q}_{CDW,2}$ on cooling through $T_{CDW,2}$ at all.

The relation between the high- and low-temperature CDW transition was discussed in detail based on x-ray diffraction results for TbTe$_3$ ($T_{CDW,1}$ = 330 K, $T_{CDW,1}$ = 41 K) [35], where it was observed that the superlattice peak intensity at $\boldsymbol{q}_{CDW,1}$ does not show any response on crossing $T_{CDW,2}$. Similarly, ARPES found two distinct regions on the Fermi surface of ErTe$_3$, which are gapped by the successive CDW transitions [49]. The gapped parts do not overlap and, hence, the two CDWs apparently do not compete for the same electronic states. Yet, our results in DyTe$_3$ put a new twist to the story because it is still surprising that phonons exhibiting strong Te displacements are not sensitive to the alleged gap opening in Te states at the Fermi surface of DyTe$_3$ connected by $\boldsymbol{q}_{CDW,2}$. The mechanism of the low-temperature CDW transition in RTe$_3$ needs also to be understood in order to explain the interesting phase diagrams, e.g., under pressure [9]. In DyTe$_3$, $T_{CDW,1}$ decreases under pressure while $T_{CDW,2}$ first increases until it hits the $T_{CDW,1}$(p) line close to 1.5 GPa. At higher pressures only one transition temperature is reported, which decreases for further increasing pressures. It is assumed that both CDW distortions are present at low temperatures and p ≥ 1.5 GPa but detailed x-ray diffraction data are still missing. Superconductivity has been found as well in DyTe$_3$ for $p \geq 1.2$ GPa [9], with a partial maximum of $T_c \approx 1.5$ K near $p = 3$ GPa where the CDW phase transition line is suppressed to zero temperature. For TbTe$_3$ and GdTe$_3$, a sharp increase of $T_c$ to $3-4$ K occurs above $4$ GPa, which was suggested to originate from either a loss of broken spatial symmetry at an orthorhombic-to-tetragonal structural phase transition intrinsic to the RTe$_3$ system, or by percolation of superconducting tellurium impurities above 5 GPa. While the detailed $T-p$ phase diagram of RTe$_3$ is not fully determined, we have shown in a work concerning the CDW material TiSe$_2$ and Cu$_x$TiSe$_2$ that a phase diagram featuring a superconducting phase emerging around the critical point where the CDW order is suppressed to zero temperature can be explained by electron-phonon coupling when the CDW soft phonon mode is also responsible for mediating superconductivity [34].

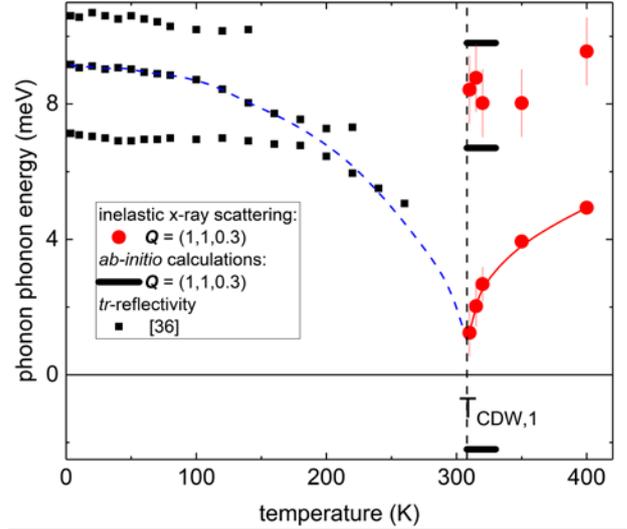

**FIG. 9.** Comparison of results from time-resolved pump-probe reflectivity measurements (squares) [36] and inelastic x-ray scattering (dots/circles). Horizontal thick bars denote the calculated phonon energies having the same symmetry as the soft mode. Negative values correspond to imaginary phonon energies [see Fig. 5(a)]. The thin solid (red) line is the soft mode temperature dependence at $T > T_{CDW,1}$. The thin dashed (blue) line is a guide to the eye for the likely amplitude/soft mode temperature dependence for $T < T_{CDW,1}$. The vertical dashed line indicates $T_{CDW,1}$ = 308 K.

## VI.    Conclusion

We have reported a combined inelastic x-ray scattering and density-functional-perturbation theory investigation of the lattice dynamics in the CDW materials DyTe$_3$. The focus of this work has been on the prediction of multiple competing soft phonon modes at the high-temperature CDW transition at $T_{CDW,1}$ = 308 K and on the evolution of these strong coupling phonon modes on cooling to below the second CDW transition at $T_{CDW,2}$ = 68 K. While the former prediction of DFPT has been verified by experiment, we find no soft mode behavior in the corresponding phonon modes on approach to $T_{CDW,2}$. While we argue that we investigated promising wavevectors based on the observation of strong CDW superlattice peaks rising below $T_{CDW,2}$, further experiments that provide comprehensive coverage of large momentum transfer space (albeit at the cost of not providing energy discrimination), e.g. thermal diffuse scattering, would be welcome to solve the puzzle on the mechanism of the low-temperature CDW transition in RTe$_3$.

**Acknowledgements:**

M.M., D.A.Z. and F.W. were supported by the Helmholtz Society under contract VH-NG-840. S.R. was supported by the Materials Sciences and Engineering Division, Office of Basic Energy Sciences, U.S. Department of Energy. Work at Stanford University was supported by the Department of Energy, Office of Basic Energy Sciences, under Contract No. DE-AC02-76SF00515. P.W. was partially supported by the Gordon and Betty Moore Foundations EPiQS Initiative through grant GBMF4414. This research used resources of the Advanced Photon Source, a U.S. Department of Energy (DOE) Office of Science User Facility operated for the DOE Office of Science by Argonne National Laboratory under Contract No. DE-AC02-06CH11357.